\pdfoutput=1

\documentclass[runningheads]{llncs}
\usepackage[T1]{fontenc}
\usepackage{graphicx}
\usepackage{amssymb}
\usepackage{amsmath}

\begin{document}
\title{Latent Spaces Enable Transformer-Based Dose Prediction in Complex Radiotherapy Plans}
\titlerunning{Latent Dose Prediction with Transformer}
% If the paper title is too long for the running head, you can set
% an abbreviated paper title here
%
\author{Edward Wang\inst{1} \and
Ryan Au\inst{1} \and
Pencilla Lang\inst{2} \and
Sarah A. Mattonen\inst{1}}
\authorrunning{Wang et al.}
% First names are abbreviated in the running head.
% If there are more than two authors, 'et al.' is used.
%
\institute{Western University \\
    \email{\{ewang225,rau23,sarah.mattonen\}@uwo.ca}\\ \and London Health Sciences Centre \\ \email{pencilla.lang@lhsc.on.ca}
}
\maketitle              % typeset the header of the contribution
\begin{abstract} %100 to 250 words
Evidence is accumulating in favour of using stereotactic ablative body radiotherapy (SABR) to treat multiple cancer lesions in the lung. Multi-lesion lung SABR plans are complex and require significant resources to create. In this work, we propose a novel two-stage latent transformer framework (LDFormer) for dose prediction of lung SABR plans with varying numbers of lesions. In the first stage, patient anatomical information and the dose distribution are encoded into a latent space. In the second stage, a transformer learns to predict the dose latent from the anatomical latents. Causal attention is modified to adapt to different numbers of lesions. LDFormer outperforms a state-of-the-art generative adversarial network on dose conformality in and around lesions, and the performance gap widens when considering overlapping lesions. LDFormer generates predictions of 3-D dose distributions in under 30s on consumer hardware, and has the potential to assist physicians with clinical decision making, reduce resource costs, and accelerate treatment planning.

\end{abstract}
\section{Introduction}
Radiation therapy (RT) is a mainstay of cancer treatment. Approximately 50\% of cancer patients worldwide will require RT over the course of their disease, although due to infrastructure and resource constraints, many patients lack access to this effective treatment \cite{Abdel-Wahab2021JCO}. The central challenge in RT is delivering sufficient radiation to treat disease while minimizing radiation toxicity. A comprehensive treatment planning and quality assurance pipeline is necessary to facilitate safe and effective treatment \cite{Fraass1998MedPhys}. Stereotactic ablative body radiotherapy (SABR) is a treatment involving the delivery of very high and conformal doses of radiation to the tumour that preserves nearby healthy organs at risk (OARs) \cite{martin2010stereotactic}. SABR is being increasingly used to treat multiple cancer lesions simultaneously, including in the lungs \cite{palmaSABRCOMET,tsai2022consolidativeCURB}. However, creating a single multi-lesion lung SABR plan is a laborious and time-consuming process taking on average 7.5 hours at our institution. Although there are many variables affecting multi-lesion SABR, including what dose to deliver to each lesion, or how many lesions to treat, planning resource constraints prevent radiation oncologists (ROs) from comparing different treatment options. The purpose of this study is to create a tool for real-time (<60s) prediction of multi-lesion lung SABR dose distributions, thereby allowing ROs to compare and select the optimal radiation prescription. 

Most existing literature on RT dose prediction focuses on single lesions \cite{kearney2020dosegan,SONG2020DeepLabs,barragan2019}. In 2020, Babier et al. hosted the OpenKBP challenge, and released a dataset of head and neck RT plans that included multiple planning targets \cite{openKBP}. However, to our knowledge, besides our own previous work \cite{wang2023GAN}, there has not been any research into the multi-lesion lung domain. Planning multi-lesion lung treatments are challenging due to the heterogeneity in the size, shape, location and number of metastatic lesions. Lesions can be treated with a wide array of prescriptions, and potential interactions between radiation delivered to nearby lesions must be accounted for. Additionally, lesions may even overlap others, such as in the case of multiple close lesions or retreatment for recurrence

Transformers are a family of autoregressive sequence prediction models, originally developed for natural language tasks, that rely on the attention mechanism \cite{vaswani2017transformer}. Through attention, transformers learn which sections of the input sequence should be more heavily weighted when predicting the next token, allowing them to capture complex relationships in the input data \cite{esser2020taming,vaswani2017transformer}. We hypothesize that this property will allow them to better account for the dose interactions between multiple lesions. To utilize a transformer for dose prediction, spatial image data must be first encoded into sequences, and then decoded back into images. Existing implementations of transformers for dose prediction \cite{JIAO2023TransDose,Wen2023Transformer,Hu2023TrDosePred} sandwich the transformer components between encoding and decoding components. The entire network is trained end-to-end, which prevents the network from adapting to variable sequence lengths and therefore varying numbers of lesions.

In this work we develop a novel Latent Dose transFormer (LDFormer) framework for RT dose prediction that operates on latent representations of patient anatomy. LDFormer fully decouples image-to-sequence encoding/decoding from sequence prediction, allowing the model to adapt to multiple lesions. We validate LDFormer on a large collection of multi-lesion lung SABR plans, and compare it to a state-of-the-art (SOTA) generative adversarial network (GAN) \cite{wang2023GAN}.

\section{Methods}
\subsection{Data}
 This study was approved by our institutional ethics review board. The dataset contains treatment plans of patients who were treated with SABR to 2-5 lung lesions (metastases $\pm$ primary) from 2010 to 2023 at a single tertiary academic centre. Patients were excluded if they received non-SABR thoracic RT prior to or in between SABR treatments. All patients received 4-D CT simulation via a Canon Aquilion LB scanner or a Philips BigBore CT scanner, with motion management performed by free breathing, gating, or deep inspiration breath hold. Treatment planning was performed with Pinnacle3 (Version 9.10, Philips Canada) or RayStation (Version 7.0, RaySearch Laboratories). Each treatment plan consists of the planning CT scan, contours of the OARs, planning target volumes (PTVs) and internal gross tumour volumes (IGTVs), and the delivered dose distribution. An IGTV is the region of space that a lesion moves through during respiratory motion, and is contoured by the treating RO. PTVs are 5 mm expansions of IGTVs, and are the regions that radiation is prescribed to. The OARs (lungs, heart, esophagus, chest wall, great vessels, and airways) were automatically contoured using Limbus Contour (Version 1.7.0, Limbus AI). Multiple plans were collected from a single patient if they received >1 multi-lesion lung SABR treatment. For example, if a patient was initially treated to three lesions, and then received additional treatment to a single lesion (new metastasis or retreatment) one year later, both a three-lesion and four-lesion plan would be collected. To combine serial treatments, the earlier treatment was non-rigidly registered to the later treatment in MIM (Version 7.2.8, MIM Software Inc) based on the planning CT scans. The PTVs and IGTVs were transferred over to the later plan, and the doses of the two treatments were combined. There were 20 dose-fractionation schemes used across all PTVs. To account for the heterogeneity in fractionations, all doses were converted to their equivalent dose in 2 Gray (Gy) fractions (EQD2) using the linear quadratic model (\(\frac{\alpha}{\beta}\)=3) \cite{McMahon_2019}. Plans were randomly divided into training ($\sim$70\%), validation ($\sim$15\%) and testing ($\sim$15\%) sets, stratified by the number of lesions. Patients with multiple treatment plans were confined to a single set. Plan characteristics are shown in Table \ref{tab:plan_char}.

\begin{table}[ht]
\centering
\caption{Plan Characteristics}
\label{tab:plan_char}
\begin{tabular}{lrrrr}
\hline
                                  & \textbf{Total}    &\textbf{Training} & \textbf{Validation} & \textbf{Testing} \\ 
\hline
Number of Plans                & 234               & 171               & 32                  & 31               \\
Number of Patients           & 198        & 157         & 23            & 18                            \\
Total Number of Lesions           & 611               & 406               &  106                & 99\\
\multicolumn{1}{r}{2 Lesions}             & 145               & 129               & 7                   & 9                \\
\multicolumn{1}{r}{3 Lesions}             & 51                & 27                & 13                  & 11               \\
\multicolumn{1}{r}{4 Lesions}             & 22                & 8                 & 7                   & 7                \\
\multicolumn{1}{r}{5 Lesions}             & 16                & 7                 & 5                   & 4                \\
Total PTV Vol. (cc) -  Median[Range]                 & 43[10-226]         & 43[11-169]         & 39[13-152]           & 52[10-226]                   \\
Number of Overlapping PTVs     & 62 & 27 & 19 & 16 \\

Prescriptions           &                   &                   &                  \\
\multicolumn{1}{r}{60Gy in 8 Fractions}         & 241               & 172              & 36                & 33               \\
\multicolumn{1}{r}{55Gy in 5 Fractions}         & 127               & 97               & 16                & 14                \\
\multicolumn{1}{r}{35Gy in 5 Fractions}         & 50                & 21               & 12                & 17               \\
\multicolumn{1}{r}{54Gy in 3 Fractions}         & 49                & 44               & 2                 & 3                \\
\multicolumn{1}{r}{30Gy in 5 Fractions}         & 33                & 17               & 8                 & 8                \\
\multicolumn{1}{r}{Other}             & 108                & 52                 & 32                  & 24                \\
\hline
\end{tabular}
\end{table}

\subsection{Data preprocessing}
OARs were voxelized to a single volume with each OAR represented by an integer and each PTV was voxelized to a separate binary volume. Voxelization was performed with the \textit{rt-utils} Python package \cite{shrestha2020rtutils}. Following our previous work \cite{wang2023GAN}, we created an initial dose estimate (IDE) based on exponential dose decay \cite{narayanasamy2018dose} to help condition the transformer. All volumes were resampled to 3mm\textsuperscript{3} spacing (linear interpolation for dose volumes, and nearest-neighbour interpolation for OAR and PTV volumes) and center cropped to 96x128x128 voxels around the lungs. The 96 voxels are in the superior-inferior direction.

\begin{figure}[htbp]
  \centering
  \includegraphics[width=\textwidth]{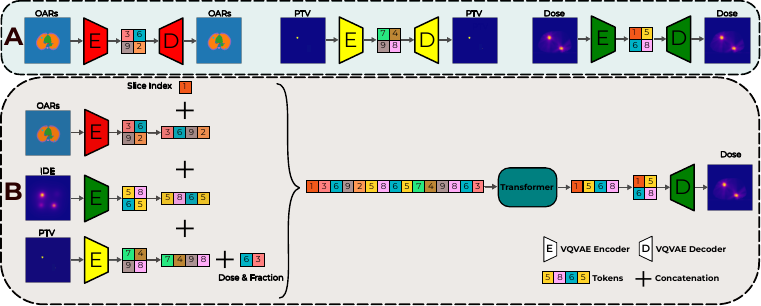}
  \caption{
    The overall workflow is shown. \textbf{A:} Vector-quantized variational autoencoders (VQVAEs) are trained to encode organs at risk (OARs), planning target volumes (PTVs), and dose into latent representations (LRs). \textbf{B:} The transformer is trained to predict the dose LR from LRs of the OARs, initial dose estimate and PTVs concatenated with the slice index and prescription. The dose LR is then decoded into a dose distribution. For simplicity, LRs are depicted as 2x2, and only one PTV is shown.
  }
  \label{workflow}
\end{figure}

\subsection{Encoding spatial data into sequences}
Prior to training the transformer, it was necessary to first encode the volumetric data describing patient anatomy and dose distributions into sequences of integer tokens (Figure \ref{workflow}A). We followed the work in \cite{esser2020taming,yan2021videogpt} and used a vector-quantized variational autoencoder (VQVAE) for this task. VQVAEs are a variant of autoencoders that map their input into a discrete latent space \cite{VQVAE}. A visual representation of VQVAEs is provided in Figure S1. For 3-D spatial data, the encoder \(E\) of the VQVAE compresses the input data \(x \in \mathbb{R}^{L \times W \times H \times c}\) into a learned latent representation of vectors \(z_v \in \mathbb{R}^{l \times w \times h \times n_z}\). Then, in the vector quantization step, the vectors in $z_v$ are replaced by their nearest vectors in a learned codebook \(Z \in \mathbb{R}^{K\times n_z}\) to form $z_v^q$. The decoder \(D\) of the VQVAE uses $z_v^q$ to create reconstruction \(\hat{x}\). The VQVAE loss function \cite{VQVAE} is
\begin{equation}
\mathcal{L}_{VQVAE} = {L}_{Rec}(x, \hat{x}) +  \lambda\| sg[z_v^q] - E(x) \|^2 + \| z_v^q - sg[E(x)] \|^2.
\end{equation}

\noindent \({L}_{Rec}\) is the reconstruction error between $x$ and $\hat{x}$, and varies per VQVAE. It is computed by mean squared error, binary cross entropy and categorical cross entropy for the dose, PTV and OAR cases respectively. $\lambda$ is a weighting factor set to 2 in this work. \textit{sg} represents the stop gradient operator which disables gradient backpropagation \cite{dhariwal2020jukebox,VQVAE}. \(Z\) is updated via exponential moving average. Further theoretical details about VQVAEs can be found in the original paper \cite{VQVAE}. During training of the OAR and dose VQVAEs, the input data was augmented by flipping across the vertical and horizontal axes with 50\% probability.

Using the VQVAEs, we encoded the input spatial data into integer sequences $s$ by replacing each vector in \(z_v^q\) with its corresponding index in \(Z\), and flattening the result into \(s \in \{0, 1, 2, \ldots, K-1\}^{N}\) where \(N=l \times w \times h\). The 2-D VQVAE formulation simply excludes the height dimension. Hyperparameters and shapes of latent representations (\(l \times w \times h\)) are provided in Table S1. We trained a 3-D VQVAE for the PTV masks, and 2-D VQVAEs for the OAR maps and dose distributions (sliced axially). Encoding PTVs in 3-D allows every sequence to contain information on the location of all PTVs. VQVAEs were used to encode the sequences \(s_{ptv}\), \(s_{oars}\), \(s_{ide}\) and \(s_{dose}\). Both \(s_{dose}\) and \(s_{ide}\) were encoded using the dose VQVAE. \(s_{ptv}\) was modified by appending the prescribed dose and fraction to the end, where dose is represented as an index in a lookup table. All sequences were concatenated to form a combined sequence \(s_{c} = \{s_{ax}, s_{oars}, s_{ptv1}, s_{ptv2}, s_{ptv3}, s_{ptv4},  s_{ptv5}, s_{dose}\}\) to feed into the transformer. \(s_{ax}\) is the axial index of the 2-D slice. The lengths of \(s_{oars}\), \(s_{ide}\) and \(s_{dose}\) are 100. The lengths of \(s_{ptv1-5}\) are 14. The total length of \(s_{c}\) is 371. We increment all values in \(s_{c}\) by 1 to reserve 0 as the padding token, for empty PTV sequences. 

\subsection{Sequence prediction with transformers}
We adapted the decoder stack from the seminal transformer paper \cite{vaswani2017transformer} to predict \(s_{dose}\) (Figure \ref{workflow}B). The decoder-only transformer generates new tokens in an autoregressive manner, in which the probability of the next token in the dose sequence \(s_{\text{dose}, i}\) depends on all previous dose tokens \(s_{\text{dose}<i}\), as well as conditioning sequences \(s_{ax}\), \(s_{oars}\), \(s_{ide}\) and \(s_{ptv1-5}\). The objective is to maximize the likelihood of $p(s_{\text{dose}, i})$, and therefore the transformer loss is the negative log-likelihood 

\begin{equation}
    L_{TF} = -\prod_{i} \log p(s_{\text{dose}, i} | s_{\text{dose}<i}, s_{\text{ax}}, s_{\text{oars}}, s_{\text{ide}}, s_{\text{ptv1-5}}).
\end{equation}

\noindent \(L_{TF}\) is only computed over the positions corresponding to tokens in \(s_{dose}\). To account for input sequences that have varying numbers of PTVs, we extended causal attention masking \cite{vaswani2017transformer} to also mask out positions of empty PTVs based on the padding token. Token position was encoded sinusoidally \cite{vaswani2017transformer}. We utilized a transformer with 4 layers, 2 heads and an embedding dimension of 128. The full model configuration is presented in Table S2. During training, we augmented the data by creating sequences from spatial data flipped along the sagittal, coronal and both planes as well as randomly selecting two permutations of PTV ordering in $s_c$, as PTV order is arbitrary. Greedy sampling was performed by simply choosing the most likely token for \(s_{\text{dose}, i}\), therefore allowing for reproducible predictions. We used LDFormer to generate dose sequences for every axial slice, which were decoded to 2-D dose slices using the decoder of the dose VQVAE. Then, all 2-D dose slices were stacked to form a 3-D distribution.

\subsection{Implementation details}
Models were implemented in PyTorch 2.0.1 (Python 3.9). Hyperparameters were tuned via grid search on the validation set. Models were trained with the AdamW optimizer \cite{adamw}. VQVAEs were trained with a learning rate of $3\times10^{-4}$. The learning rate for the transformer consisted of a linear warmup for 200 epochs to $1\times10^{-4}$ followed by cosine decay \cite{radford2018improving}. The batch sizes for the 3-D VQVAE, 2-D VQVAEs, and the transformer were 16, 512 and 512 respectively. The 3-D VQVAE was trained for 1000 epochs. The 2-D VQVAEs were trained for 5000 epochs. The transformer was trained for 1000 epochs. VQVAEs were trained on a NVIDIA V100 32GB GPU with training times ranging from $\sim$4 hours to $\sim$1.5 days. The transformer was trained on a NVIDIA 3090 24GB GPU for $\sim$1 day. Training and evaluation code is available at https://github.com/edwardwang1/LDFormer. 

\subsection{Model Evaluation}
We evaluated LDFormer on the testing set by comparing the predicted dose to the ground truth dose on the basis of dose-volume-histogram (DVH) metrics of OARs, conformality metrics of the PTVs, and mean absolute difference (MAD) across all structures. All metrics were calculated in EQD2. The DVH metrics are taken from the dose constraints used in the ongoing phase III multi-lesion SABR-SYNC clinical trial \cite{SABR_SYNC}. They are the maximum dose to 5 cubic centimeters ($D_{5cc}$) of the esophagus, chest wall and airways, $D_{10cc}$ of the great vessels, $D_{15cc}$ of the heart, the volume of lung receiving less than 14 Gy ($CV\textsubscript{14}$), and the percent of lung receiving above 15 Gy ($V\textsubscript{15}$). Both $CV\textsubscript{14}$ and $V\textsubscript{15}$ are the EQD2 equivalent of the contraints from \cite{SABR_SYNC}. DVH metrics were calculated from the masks of the OARs minus the IGTVs. The conformality metrics used are the heterogeneity index ($HI$), and the maximum dose at 1 cm and 2 cm away from the PTV ($D_{1cm}$, $D_{2cm}$). $HI$ is the ratio of the maximum dose inside the PTV to the prescription dose \cite{kearney2020dosegan,feuvret2006conformity}. As dose is not guaranteed to decrease with increasing distance from the PTV due to other lesions, $D_{1cm}$ and $D_{2cm}$ were calculated from a 1 voxel ($27mm^3$) thick sphere at 1 cm and 2 cm from the PTV. $HI$, $D_{1cm}$, and $D_{2cm}$ were calculated for all lesions, as well as only lesions with overlap. Finally, we calculated the MAD between the predictions and ground truth across all OARs and PTVs. We compared LDFormer to our previous implementation of a GAN as described in \cite{wang2023GAN} on all metrics. Significance testing was performed in Python 3.9 using the T-test for normal data and Wilcoxon rank-sum test for non-normal data, with normality assessed by the Shapiro-Wilk test.

\begin{table}[ht]
\centering
\caption{The absolute differences in the dose-volume-histogram and conformality metrics between predicted doses and ground truth in the testing set are reported for LDFormer and the GAN as mean$\pm$SD. Conformality metrics are calculated over all lesions, as well as only lesions with overlap (Ov). Bold font indicates significantly better performance (p<0.05). The unit of CV\textsubscript{14} is cc. The unit of V\textsubscript{15} is \%. The units of D\textsubscript{Xcc} and D\textsubscript{Xcm} are EQD2 Gy with $\frac{\alpha}{\beta}=3$. HI is dimensionless. Ln=Lung, Es=Esophagus, Hr=Heart, Aw=Airways, Gv=Great Vessels, Cw=Chest Wall.}
\label{tab:quant_results}
\begin{tabular}{lccccccc}
\hline
\textbf{Model} & \textbf{LnCV\textsubscript{14}} & \textbf{LnV\textsubscript{15}} & \textbf{EsD\textsubscript{5cc}} & \textbf{HrD\textsubscript{15cc}} & \textbf{AwD\textsubscript{5cc}} & \textbf{GvD\textsubscript{10cc}} & \textbf{CwD\textsubscript{5cc}} \\
\hline
LDFormer & 122$\pm$120 & 3.3$\pm$4.1 & 4.5$\pm$8.2 & 5.8$\pm$8.8 & 5.8$\pm$7.1  & 5.3$\pm$4.5 & 22$\pm$20 \\
GAN & 70$\pm$53 & 1.7$\pm$1.5 & 4.7$\pm$6.5 & 5.3$\pm$7.2 & 3.6$\pm$5.1 & 4.4$\pm$4.8 & 17$\pm$22\\
\hline
\textbf{Model} & \textbf{HI} & \textbf{D\textsubscript{1cm}} & \textbf{D\textsubscript{2cm}} & \textbf{OvHI} & \textbf{OvD\textsubscript{1cm}} & \textbf{OvD\textsubscript{2cm}}\\
\hline
LDFormer & \textbf{0.44$\pm$0.40} & \textbf{32$\pm$28} & 22$\pm$19 & \textbf{0.67$\pm$0.49} & \textbf{44$\pm$32} & \textbf{39$\pm$25} \\
GAN & 0.74$\pm$0.69 & 53$\pm$40 & 27$\pm$28 & 1.71$\pm$0.66 & 97$\pm$35 & 65$\pm$40\\
\hline
\end{tabular}
\end{table}

\section{Results}
Quantitative results are shown in Table \ref{tab:quant_results}. Across all PTVs, LDFormer significantly outperformed the GAN on $HI$ and $D_{1cm}$. There were 16 lesions in the testing set that overlapped with other lesions. For these, LDFormer performed significantly better on all PTV conformality metrics and the magnitude of improvement was greater, as the GAN tended to overestimate dose due to overlap. Across all DVH and MAD metrics, LDFormer performed similarly to the GAN. MAD summary results are provided in Table S3. Figure \Ref{fig:qualResults} shows LDFormer and GAN predictions for two testing set plans with overlapping lesions. In the GAN dose, hotspots can be seen inside overlapping lesions which are not present on the LDFormer dose. Minor step artifact is visible on the LDFormer dose, caused by the stacking of 2D predictions. The inference time for a full 3-D dose distribution of a 5-lesion plan, including loading weights, is 8.5s for the GAN, and 28.7s for LDFormer on an NVIDIA 3090 24GB GPU.

\begin{figure}[htbp]
  \centering
  \includegraphics[width=1.0\textwidth]{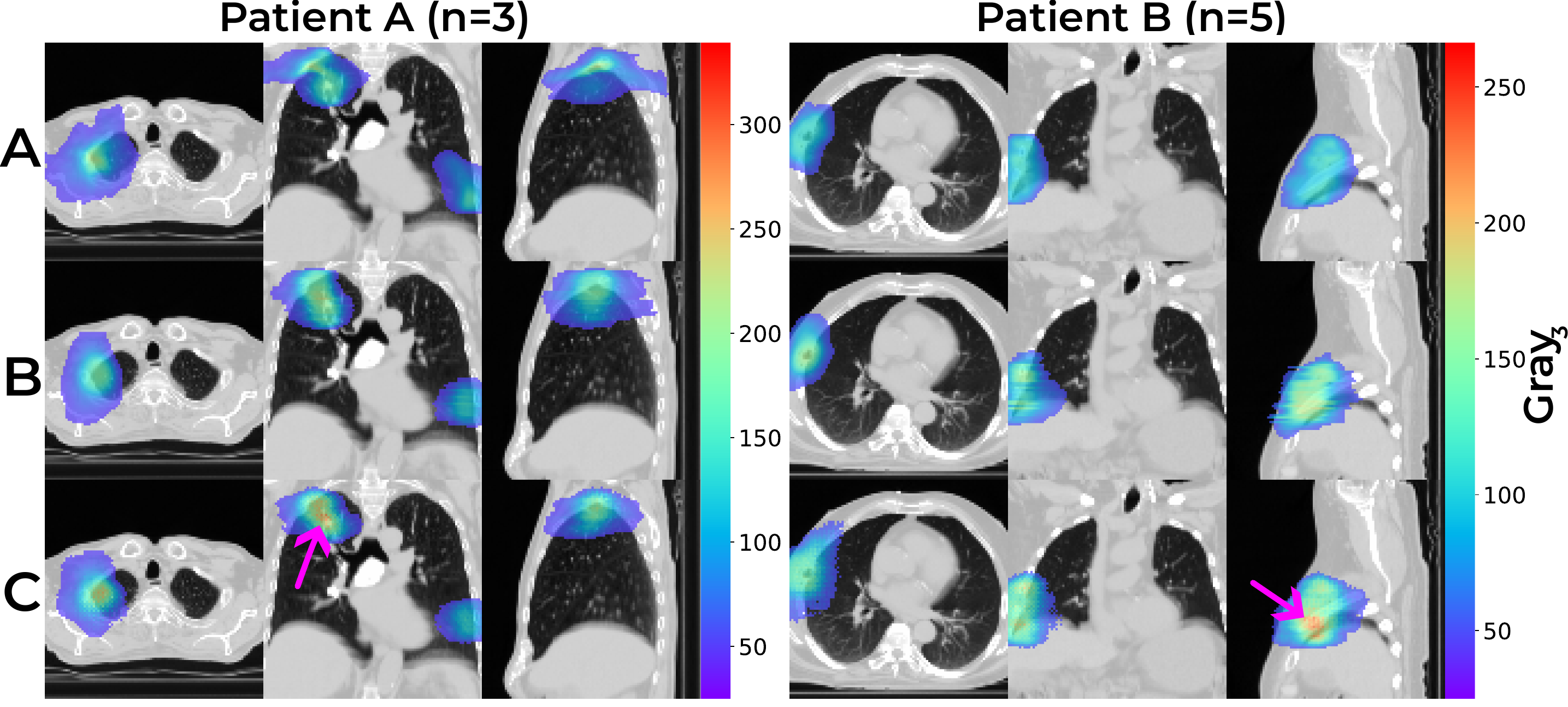}
  \caption{Axial, coronal, and sagittal views of the (A) ground truth, (B) LDFormer and (C) GAN doses are shown for two testing set patients with overlapping lesions. Arrows indicate hotspots in overlapping lesions. The unit of the colourbar is EQD2 Gy ($\frac{\alpha}{\beta}=3$).} 
  \label{fig:qualResults}
\end{figure}

\section{Discussion}
Evidence is growing in favour of treating more metastases \cite{palmaSABRCOMET,tsai2022consolidativeCURB} with SABR, increasing resource requirements. For patients with multiple lesions, there are many permutations of which lesions are treated, and the radiation prescribed to each lesion. A real-time dose prediction tool would allow ROs to quickly compare potential treatments both visually and quantitatively to select the best one, rather than monopolizing resources to create a full plan for every option. The predicted dose distributions can also be used as optimization targets during inverse planning \cite{babier2022openkbpOpt,fan2019automaticTP,mcintosh2017fullyAutomatedTP} further accelerating planning. In this work, we leverage the powerful modelling capabilities of the transformer \cite{vaswani2017transformer} for this task, and build a novel framework capable of making accurate predictions in under 30s.

LDFormer outperforms our previous GAN approach \cite{wang2023GAN} on PTV conformality metrics, and is competitive with the GAN on DVH metrics. Overlap analysis demonstrates that the greatest advantage of LDFormer is in dose prediction for plans with overlapping PTVs, supporting the hypothesis that attention allows LDFormer to better account for inter-lesion dose interactions. Such cases are most common in the context of retreatment, but may also occur when multiple lesions are grouped close together. It is challenging for ROs to prescribe an appropriate treatment for these patients, and it is in these complex cases that they would benefit most from dose prediction tools like LDFormer. 

A limitation of this work is the modest size of the dataset. LDFormer consists of a small transformer operating on heavily compressed latent representations of dose and anatomical data. Additional training data would enable a larger model and longer sequence lengths (i.e. larger, less compressed latent representations), reducing error introduced by the VQVAEs. Further, transformers lack inductive spatial biases \cite{esser2020taming} present in convolutional neural networks (such as the GAN used for comparison \cite{wang2023GAN}) and therefore require larger datasets for image-based tasks \cite{visionTransformer}. Another limitation common to dose prediction methods is that the output is not guaranteed to be physically deliverable, even if the model was exclusively trained on real plans. However, the real benefit of these dose prediction techniques is the reduction of required resources, not that the patient's real treatment exactly matches the prediction. An "undeliverable" prediction may still be beneficial to RO clinical decision making and during inverse planning. To study this, we are preparing a prospective validation series to quantify the resource savings of introducing LDFormer into the clinical workflow at our centre. 

\section{Conclusion}
In this work, we present the LDFormer framework for multi-lesion lung SABR dose prediction. LDFormer outperforms a SOTA GAN in PTV conformality metrics, and is most beneficial for plans with overlapping lesions. Multi-lesion lung SABR is an effective but resource intensive treatment. Our work has the potential to reduce resource burden and increase the adoption of this technique.

\begin{credits}
\subsubsection{\ackname} This project was funded by the Gerald C. Baines Foundation, donor support through the London Health Sciences Foundation, the Keith Samitt Translational Cancer Research Grant, and trainee support from the Canadian Institutes of Health Research and the Natural Sciences and Engineering Research Council of Canada. The authors would like to thank Karen Eddy for assisting with the clinical data curation.
\end{credits}
\newpage
%
% ---- Bibliography ----
%
% BibTeX users should specify bibliography style 'splncs04'.
% References will then be sorted and formatted in the correct style.
%
\bibliographystyle{splncs04}
\bibliography{mybibliography}

% This is samplepaper.tex, a sample chapter demonstrating the
% LLNCS macro package for Springer Computer Science proceedings;
% Version 2.21 of 2022/01/12
%
%\documentclass[runningheads]{llncs}
%
%\usepackage[T1]{fontenc}
% T1 fonts will be used to generate the final print and online PDFs,
% so please use T1 fonts in your manuscript whenever possible.
% Other font encondings may result in incorrect characters.
%
%\usepackage{graphicx}
% Used for displaying a sample figure. If possible, figure files should
% be included in EPS format.
%
% If you use the hyperref package, please uncomment the following two lines
% to display URLs in blue roman font according to Springer's eBook style:
%\usepackage{color}
%\renewcommand\UrlFont{\color{blue}\rmfamily}
%\urlstyle{rm}
%
%\begin{document}
%
%

\renewcommand\thefigure{\thesection.\arabic{figure}}    
\setcounter{figure}{0}
\renewcommand\thetable{\thesection.\arabic{table}}    
\setcounter{table}{0}    

{
\centering
\Large\bfseries
Supplementary Material for Latent Spaces Enable Transformer-Based Dose Prediction in Complex Radiotherapy Plans\par
}

\vspace{15ex}

\begin{figure}[htbp]
  \centering
  \renewcommand{\thefigure}{S\arabic{figure}}
  \includegraphics[width=1.0\textwidth]{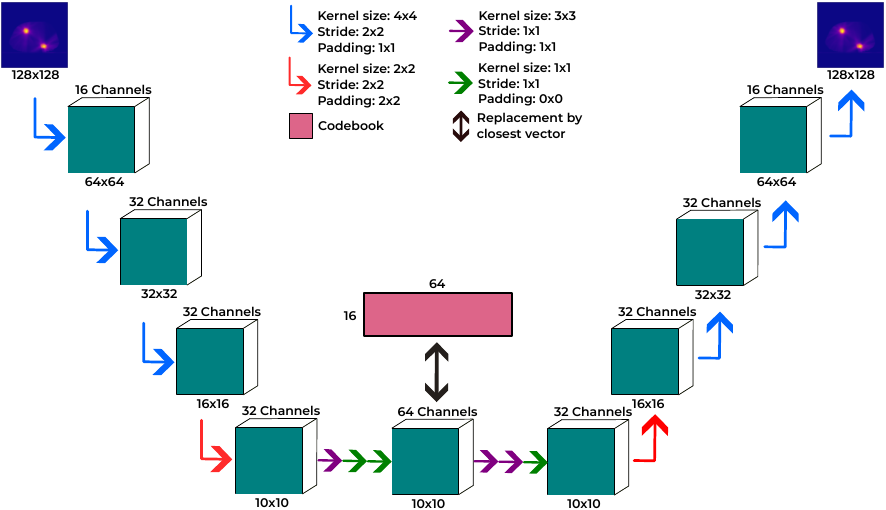}
  \caption{The architecture of the 2-D vector-quantized variational autoencoders (VQVAE) is shown. Arrows pointing to the right are convolution layers, and arrows pointing up are transpose convolution layers. The codebook is a 2-D matrix of embedding vectors. The 3-D VQVAE has 6 downsampling convolutions instead of 4, with the properties of the last downsampling convolution provided in Table S1.} 
  \label{fig:qualResults}
\end{figure}

\begin{table}[htbp]
\label{table:SupVQVAEConfig}
\centering
\renewcommand{\thetable}{S\arabic{table}}
\caption{Hyperparameters of the 3-D and 2-D vector-quantized variational autoencoders are shown, along with input and latent space dimensions. The last column displays the searched hyperparameter configurations. DS = Downsampling}
\begin{tabular}{llll}
\hline
                             & \textbf{3-D} & \textbf{2-D} & \textbf{HP Search}\\ 
\hline
Input Dimensions             & 96x128x128         & 128x128           &- \\
Latent Space Dimensions            & 3x2x2              & 10x10              &-\\
Sequence Length              & 12                 & 100                &-\\
Number of DS Convolutions     & 6                  & 4                  &-\\
Last DS Convolution Kernel Size & 1x4x4              & 2x2               &- \\
Last DS Convolution Stride      & 1x2x2              & 2x2                &-\\
Last DS Convolution Padding     & 0x1x1              & 2x2                &- \\
Convolution Channels       & 4,8,16,32,32,32                    & 16,32,32,32         & [16, 32, 64]  \\
Number of Embedding Vectors ($K$)               & 64                 & 64                & [32, 64, 128]    \\
Dimensions of Embedding Vectors ($n_z$)         & 16                 & 16                 & [16, 32] \\
\hline
\end{tabular}
\end{table}

\begin{table}[]
\centering
\renewcommand{\thetable}{S\arabic{table}}
\caption{Hyperparameters of transformer model are shown. The last column displays the searched hyperparameter configurations. The vocabulary size is the number of axial slices plus 1 for padding, and block size is equal to sequence length minus 1.}
\begin{tabular}{l@{\hspace{1em}}l@{\hspace{1em}}l}
\hline
\textbf{Parameter}                 & \textbf{Value} & \textbf{HP Search}\\ 
\hline
Number of Layers          & 4            & [2, 3, 4]\\
Overall Model Dimension   & 128            &-\\
Number of Attention Heads & 2              &-\\
Attention Head Dimension  & 64             &-\\
Vocabulary Size           & 97             &-\\
Block Size                & 370           &-\\
\hline
\end{tabular}
\end{table}

% {
% \large
% \bfseries
% \noindent
% Mean Absolute Difference Results\par
% }

\begin{table}[htbp]
\centering
\renewcommand{\thetable}{S\arabic{table}}
\caption{The mean absolute difference (MAD) between the predicted and ground truth dose distributions in the test set are reported for LDFormer and the GAN as mean$\pm$SD (Gray). There was no significant difference in MAD between the two models across any organs at risk or the planning target volumes (PTVs). Gv=Great Vessels, Cw=Chest Wall}
\label{tab:quant_results}
\begin{tabular}{lccccccc}
\hline
\textbf{Model} & \textbf{Lung} & \textbf{Esophagus} & \textbf{Heart} & \textbf{Airways} & \textbf{Gv} & \textbf{Cw}  & \textbf{PTV}\\
\hline
LDFormer & 3.7$\pm$1.9 & 2.4$\pm$2.5 & 1.8$\pm$2.1 & 2.9$\pm$2.5  & 2.7$\pm$2.1 & 2.1$\pm$1.0 & 33.9$\pm$13.7\\
GAN & 3.2$\pm$1.4 & 2.2$\pm$2.1 & 1.7$\pm$1.7 & 2.3$\pm$1.8 & 2.7$\pm$2.0 & 2.0$\pm$0.9 & 28.5$\pm$9.3\\
\hline
\end{tabular}
\end{table}

%\bibliographystyle{splncs04}
%\bibliography{mybibliography}

%\end{document}

\end{document}